\documentclass[pra,superscriptaddress,floatfix]{revtex4}
\usepackage{amsmath}
\usepackage{amstext}
\usepackage{latexsym}
\usepackage{graphicx}

\newcommand{\eps}{\varepsilon}
\newcommand{\abs}[1]{\vert#1\vert}

\newcommand{\aver}[1]{\langle#1\rangle}

\newcommand{\virg}[1]{``{#1}''}

\begin{document}
\title{Pairing of one-dimensional Bose-Fermi mixtures with unequal masses}

\author{Matteo Rizzi}
\affiliation{NEST-CNR-INFM and Scuola Normale Superiore, I-56126
Pisa, Italy}
\affiliation{Max Planck Institut f\"ur QuantenOptik, Hans Kopfermann Strasse
  1, D-85748 Garching, Germany}
\author{Adilet Imambekov}
\affiliation{Department of Physics, Harvard University, Cambridge
MA 02138, USA}
\affiliation{Department of Physics, Yale University, P.O. Box 208120,
New Haven, CT 06520-8120, USA}

\date{\today}
\begin{abstract}
We have considered one dimensional Bose-Fermi mixture with equal densities and
unequal masses using numerical DMRG. For the mass ratio of K-Rb mixture and
attraction  between bosons and fermions, we determined the phase diagram. For
weak boson-boson interactions, there is a direct transition between
two-component Luttinger liquid and collapsed phases as the boson-fermion
attraction is increased. For strong enough boson-boson interactions, we find
an intermediate ``paired'' phase, which is a single-component Luttinger liquid
of composite particles. We investigated correlation functions of such a
``paired'' phase, studied the stability of ``paired'' phase to density
imbalance, and discussed various experimental techniques which can be used to
detect it. 
\end{abstract}

\pacs{03.75Mn, 03.75.Hh, 73.43.Nq, 75.40.Mg}

\maketitle

\begin{figure}[tbhp]
 \begin{center}
 \includegraphics[width=0.7\linewidth]{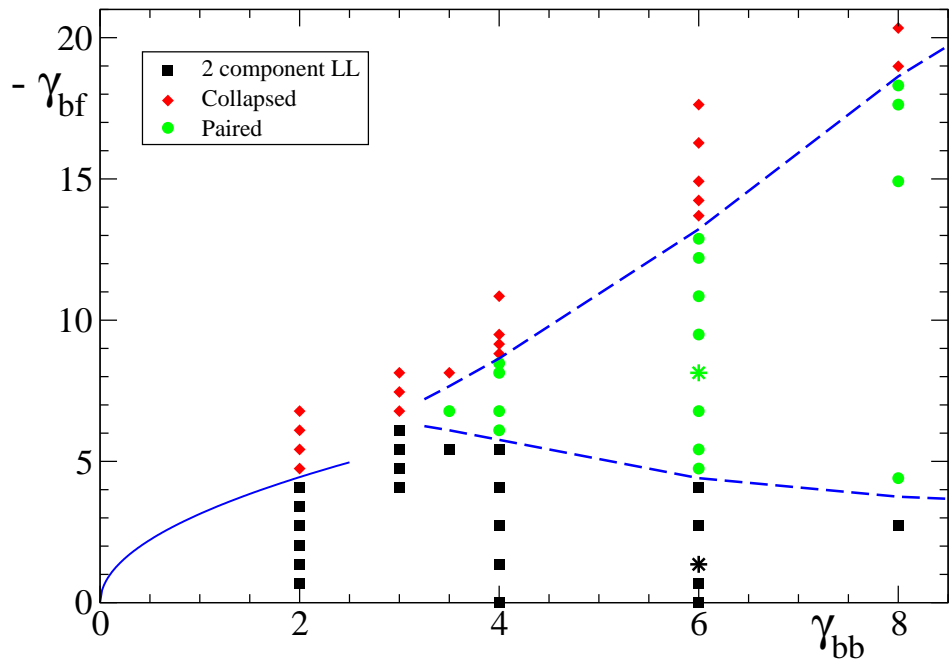}
\caption{\label{fig:phasediag} \small{Phase diagram for K-Rb mixture with equal
  density of bosons and fermions. $\gamma_{bb}$ and $\gamma_{bf}$ are defined
  by Eq. (\ref{gammadefs}), and we show here the attractive side ($\gamma_{bf}
  < 0$). Different symbols and colors identify the three different phases
  involved:
 squares are in the two-component Luttinger liquid (LL) state,
 diamonds  are in the collapsed state, 
 whereas circles stand for the paired phase we found in between.
 Sample points, considered in Figs.~\ref{fig:Pop}, \ref{fig:Corr}, and
 \ref{fig:NkCorrPaired}, are identified by stars. Solid line is the mean-field
 prediction~\cite{Das}, which should be good for small $\gamma_{bb}$ and
 matches the data for smallest $\gamma_{bb}$ available.
 Dashed lines are tentative boundaries between the different phases. Notice
 the existence of a threshold in $\gamma_{bb}$ to get the paired phase.}}
\end{center}
\end{figure}

Recent developments in the cooling and trapping of atomic gases open exciting
opportunities for experimental studies of interacting systems under
well-controlled conditions. Using Feshbach resonances~\cite{Feshbach} and/or
optical lattices~\cite{Jaksch98,Bloch} it is possible to reach strongly
interacting regimes, where correlations between atoms play a crucial role. The
effect of interactions is most prominent for low-dimensional systems, and
recent experimental realization~\cite{Weiss,Paredes} of a strongly interacting
Tonks-Girardeau (TG) gas of bosons opens new perspectives in experimental
studies of strongly interacting systems in 1D. In particular, one can
experimentally study the behavior of Bose-Fermi (BF) mixtures~\cite{bfexp} in
1D. Due to the lack of candidate systems in traditional solid state
experiments, this topic did not attract much theoretical attention until
recently. By now, properties of 1D BF mixtures have been investigated using
mean-field approximation~\cite{Das}, Luttinger liquid (LL) formalism
~\cite{CazalillaHo,Mathey,MatheyWang,Mathey_commensurate}, $T$-matrix
approximation~\cite{Tmatrix}, exact solutions~\cite{exactsolution}, and
numerical methods~\cite{jap_numerics, Sengupta05,Pollet05,Pollet06}.
The mean-field approximation is unreliable in 1D, the LL approach describes the phase
diagram in terms of universal parameters which are hard to relate to
experimentally controlled parameters, and the exact solution is restricted only to
a certain region of parameter space. Most of the numerical work so far
considered BF mixtures in optical lattices with fillings of the order of
unity. In such a regime the analysis of the phase diagram is complicated, since the
physics of the Mott transition plays an important role. In this article, we are
mainly interested in the properties of 1D mixtures without optical lattice,
for the regime of parameters directly relevant to current K-Rb experiments
~\cite{currBFM}. Our main result is summarized in Fig.~\ref{fig:phasediag}. For
K-Rb mixture with equal density of bosons and fermions, we find the
``pairing'' phase for moderate interaction strengths. Such a phase was
previously discussed in Ref.~\cite{CazalillaHo} for the case of almost equal
masses and strong interactions based on LL formalism, but its relevance for
current experiments with K-Rb mixtures has not been addressed before. We note
that the existence of ``pairing'' discussed in this article does not require the
presence of commensurate optical lattice, as in Ref.~\cite{Kuklov03}.

The general Hamiltonian	of a 1D BF mixture is given by
\begin{equation} \label{eq:initham}
\mathcal{H} = \int_0^L \left( \frac{\hbar^2}{2m_b}\partial_x
\Psi_b^\dagger\partial_x \Psi_b+\frac{\hbar^2}{2m_f}\partial_x
\Psi_f^\dagger\partial_x \Psi_f \right) \mathrm{d} x + \int_0^L
\left(\frac12g_{bb}\Psi_b^\dagger \Psi_b^\dagger\Psi_b\Psi_b + g_{bf}
\Psi_b^\dagger \Psi_f^\dagger\Psi_f\Psi_b \right) \mathrm{d} x \ ,
\end{equation}
where $\Psi_b, \Psi_f$ are boson and fermion operators, $m_b, m_f$ are the
masses, and $g_{bb}, g_{bf}$ are boson-boson and boson-fermion interaction
strengths. Well away from confinement induced resonances~\cite{Olshanii98}, 1D
interactions are given by 
$$ g_{bb} = 2 \hbar \omega_{b\perp} \, a_{bb}\;\; \mbox{and} \;\;
g_{bf}= 2 \hbar \frac{\omega_{b\perp} \omega_{f\perp} (m_b+m_f)}{\omega_{b\perp} m_b
 +\omega_{f\perp}m_f} \, a_{bf}\, ,$$
 where $\omega_{b\perp},\omega_{f\perp}$ are transverse
 confinement frequencies, and $a_{bb},a_{bf}$ are 3D scattering
 lengths. $g_{bb}$ and $g_{bf}$ can be controlled by changing transverse
 confining frequencies, or by varying scattering lengths using Feshbach
 resonances~\cite{Feshbach}. For a K-Rb mixture in the absence of magnetic
 field, $a_{bb}>$ and $a_{bf}<0,$ so in this article we will study the regime
 when boson-fermion interaction is attractive and boson-boson interaction is
 repulsive. The phases of Hamiltonian~(\ref{eq:initham}) in the most general case
 depend on four dimensionless parameters, which we choose to be $m_b/m_f$, 
 $n_b/n_f$, 
\begin{equation}
\gamma_{bb}=\frac{m_b g_{bb} }{\hbar^2 n_b}\;\; \mbox{and} \;\;
\gamma_{bf}=\frac{\sqrt{m_b m_f} g_{bf}}{\hbar^2 \sqrt{n_b n_f}}.
\label{gammadefs}
\end{equation}
 Here $n_f, n_b$ are fermion and boson densities, and $\gamma_{bb}$ and
 $\gamma_{bf}$ are dimensionless interaction parameters. Similar to
 Lieb-Liniger model~\cite{LL}, strongly interacting regime corresponds to
 $\gamma_{bb},|\gamma_{bf}|\gg 1.$ If $\gamma_{bb}<0$ (attractive bosons), the
 system is always unstable towards boson collapse. For $\gamma_{bb}>0$ the
 system can still collapse for $\gamma_{bf}<0,$ or phase separate for
 $\gamma_{bf}>0$~\cite{CazalillaHo}. If these two scenarios are not realized
 and densities of bosons and fermions are incommensurate, then from LL theory
 ~\cite{CazalillaHo,Mathey,MatheyWang} one expects a two-component LL, with
 power law decay of all correlations. If densities are commensurate, one can
 expect~\cite{CazalillaHo, Mathey_commensurate} a nontrivial pairing, resulting
 in the exponential decay of certain correlation functions  and in the opening
 of the gap. In what follows we will concentrate on the latter case for an
 experimentally relevant K-Rb mixture, so we will fix
 $n_b/n_f=1$, $m_b/m_f=87/40$ and consider $\gamma_{bf}$ negative. 

For numerical purposes we consider the discretized version of
Hamiltonian~(\ref{eq:initham})
to be an open boundary chain with unity lattice constant
and $L$ sites. Similar to the Lieb-Liniger model being a low filling fraction
limit of Bose-Hubbard model~\cite{SchmidtCazalilla},
Hamiltonian~(\ref{eq:initham})
is the low filling fraction limit of the following lattice
Hamiltonian:
\begin{equation}\label{eq:lattham}
\mathcal{H}_{L} = - \sum_{i=1}^{L-1} \left( t_b \left( b_i^{\dagger}
b_{i+1} + H.c. \right) - t_f \left( f_i^{\dagger}
f_{i+1} + H.c. \right) \right) + \sum_{i=1}^{L} \left( \frac{U_{bb}}{2}\,
b_i^{\dagger} b_{i} (b_i^{\dagger} b_{i} -1) + U_{bf}\,
b_i^{\dagger} b_{i} \; f_i^{\dagger} f_{i} \right).
\end{equation}
We note that many different lattice Hamiltonians give continuum Hamiltonian
(\ref{eq:initham}) in the low density limit, and the choice of lattice Hamiltonian
is not unique. For low fillings $\nu_b$ and $\nu_f,$ dimensionless interaction
parameters are given by
\begin{equation}
\gamma_{bb}\approx \frac{U_{bb} }{2 t_b \nu_b}\;\; \mbox{and} \;\;
\gamma_{bf}\approx \frac{U_{bf}}{2\sqrt{t_b \nu_b t_f \nu_f}}.
 \label{gammacorrespondence}
\end{equation}
Most of our simulations were performed at densities $\nu \simeq 1/4,$ but some
of the results were checked for $\nu \simeq 1/8$. The fact that we use finite
filling fractions only slightly changes Eq.~(\ref{gammacorrespondence}), but
does not affect the phase diagram qualitatively. We stress here that our
results were checked also against commensurability effects by testing them at
filling fractions $\nu = 23/96$ and $\nu = 25/96$ too. Thus we can safely say
that pairing effects are not depending on the particular $\nu$ value
chosen. The expectation values $\aver{\ldots}$ of one- and two-body
operators over  the ground state of $\mathcal{H}_{L}$ have been evaluated by
means of  the DMRG method~\cite{DMRG}, which provides a practically exact
solution for any value of the couplings and allows one to measure correlation
functions with both statistics on the equal footing. We used up to $L=96$
chains, with local dimension $d=10$ (up to 4 bosons per site) and truncation
up to $m=256$ states. Discarded probabilities amount to less than $\eps = 5 \cdot
10^{-7}$. Small values of $\gamma_{bb}$ with weak interparticle interactions,
however, are not easy to study with this method, since high occupation number
for bosons should  be taken into account. We thus resorted to mean-field
predictions in this area. 

\begin{figure}[tbhp]
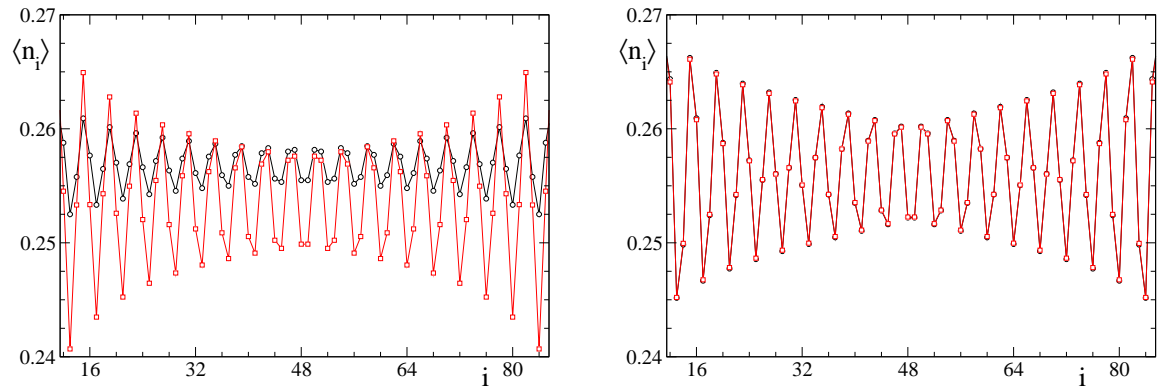

 \begin{center}
\tabcolsep=0.025\linewidth
\begin{tabular}{cc}
  \includegraphics[width=0.40\linewidth]{Pop.usual.Ubb3.0.Ubf1.0.eps} &
  \includegraphics[width=0.40\linewidth]{Pop.paired.Ubb3.0.Ubf6.0.eps}
\end{tabular}
  \caption{\label{fig:Pop} Density profiles:
    \small{black circles are bosons, red squares fermions.
      In the first panel we show results for a typical point in the usual
      mixed phase ($\gamma_{bb} = 6.0, \gamma_{bf} = -1.36$). Both the species
      are spread around the chain and no pinning is evident between
      densities. In the second panel, the \virg{locking} of densities for the
      paired phase is evident. Right panel corresponds to the values
      $\gamma_{bb} = 6.0, \gamma_{bf} = -8.14$.	 Friedel oscillation frequency
      is in both cases given, as expected, by $\omega_* = \pi \nu_*$ with
      $\nu_* \simeq 1/4$ being the species' density in the bulk of the
      system.}
}
\end{center}
\end{figure}

\begin{figure}[tbhp]
 \begin{center}
\tabcolsep=0.025\linewidth
\begin{tabular}{cc}
  \includegraphics[width=0.40\linewidth]{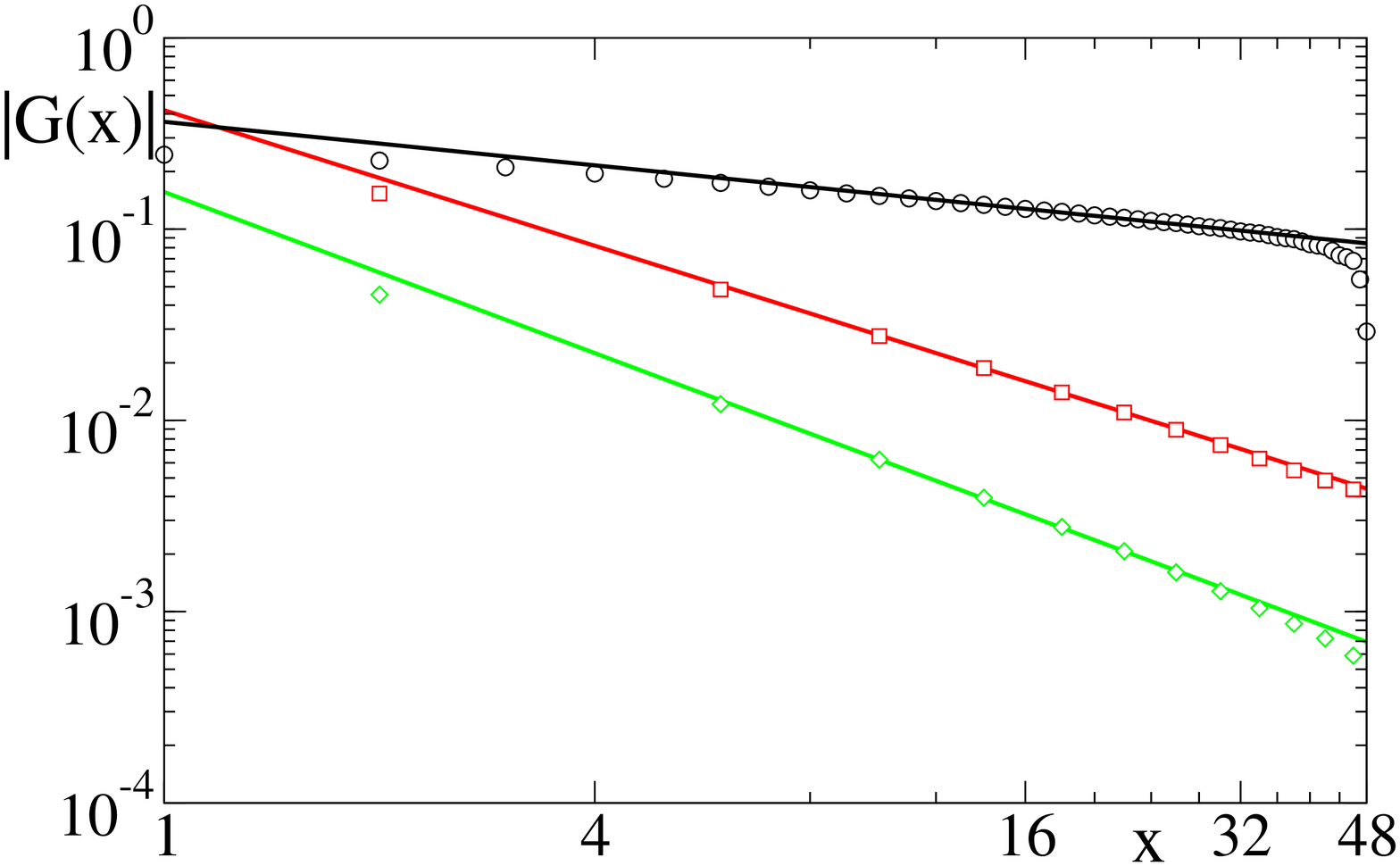} &
  \includegraphics[width=0.40\linewidth]{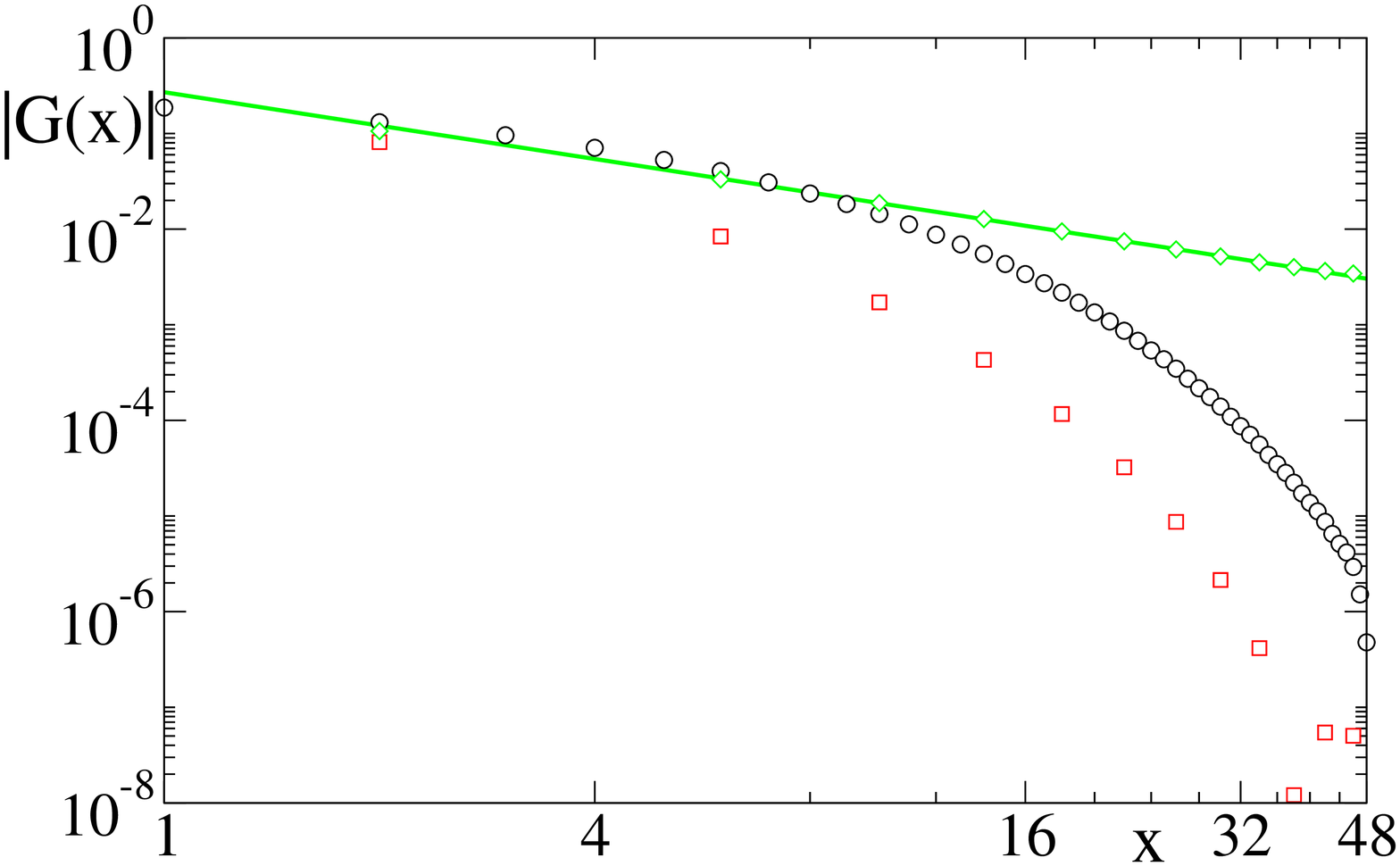}
\end{tabular}
  \caption{\label{fig:Corr}Correlation functions:
    \small{same sample points as in Fig.~\ref{fig:Pop}, same color code.
    For clarity, oscillations are not shown and only the decay of the envelope
    functions is presented. $G(x)$ means here $G(L/2, L/2+x)$. Green diamonds
    stand for composite particles created by $\Delta^{\dagger} = b^{\dagger}
    f^{\dagger}$. 
    In the first panel (two-component Luttinger liquid state)  all the three
     types of correlation functions exhibit an algebraic decay. For bosons
     $G_b(x) \propto x^{-1/(2K_b)}$ with $K_b = 1.45 \pm 0.05$, whereas for
     fermions $G_f(x) \propto x^{-1/(2K_f)-K_f/2}$ with $K_f = 0.98 \pm
     0.02$. On the other hand, for ``paired'' phase (second panel) only
     $G_{bf}$ shows an algebraic decay $\simeq x^{-1/(2K_{bf})-K_{bf}/2}$ with
     the Luttinger parameter $K_{bf} = 0.95 \pm 0.02,$ while $G_{b}$ and
     $G_{f}$ decay exponentially with distance.}
}
\end{center}
\end{figure}

\begin{figure}[tbhp]
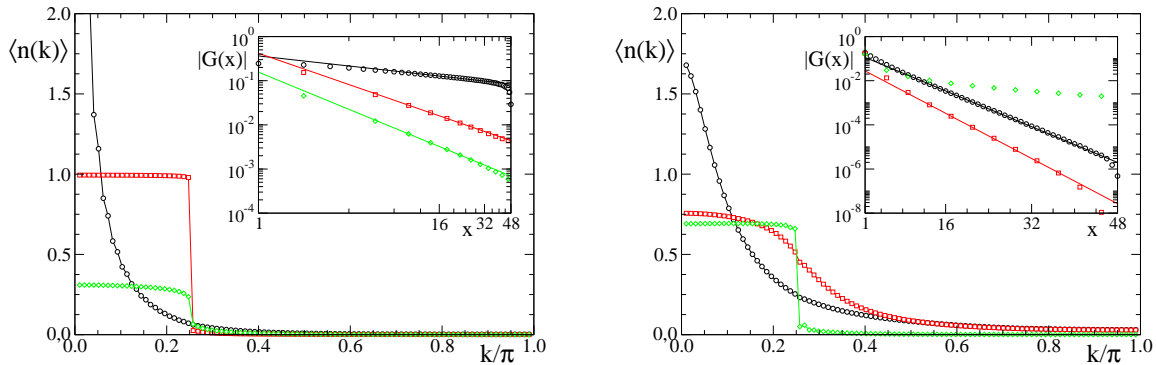

 \begin{center}
\tabcolsep=0.025\linewidth
\begin{tabular}{cc}
  \includegraphics[width=0.40\linewidth]{CorrExp.usual.Ubb3.0.Ubf1.0.eps} &
  \includegraphics[width=0.40\linewidth]{CorrExp.paired.Ubb3.0.Ubf6.0.eps}
\end{tabular}
  \caption{\label{fig:NkCorrPaired} Momentum distributions:
\small{
  same sample points and same color code of correlations as in Fig.~\ref{fig:Corr}.
  In the first panel, data for the two-component Luttinger liquid phase, where
  a Fermi surface for individual fermions and a tightly peaked distribution
  for bosons can be clearly seen. In the second panel, momentum distributions
  in the ``paired'' phase. We highlight
      the washing out of the Fermi surface for individual $^{40}$K and the
      wide broadening of the bosonic $^{87}$Rb distribution. 
      On the contrary, a sharp step-like feature in the composite particles'
      mode occupation is present around $\pi/4,$ indicating the algebraic
      decay of	  ``pairing'' correlation function $G_{bf}(x)$. Bosonic and
      fermionic particle correlations decay exponentially (see the inset,
      which is the same data as in the second panel of Fig.~\ref{fig:Corr}, but
      in log-linear scale).
}
      }
\end{center}
\end{figure}

Local density profiles were measured and are plotted in Fig.~\ref{fig:Pop} for
two sample points (corresponding to stars in Fig.~\ref{fig:phasediag}).
In our simulations we calculated the following correlation functions:
bosonic Green function $G_b (i,j)  = \aver{b_i^{\dagger} b_j},$
fermionic Green function $G_f (i,j)= \aver{f_i^{\dagger} f_j}$
and ``pairing''  correlation function $G_{bf} (i,j)
= \aver{\Delta_i^{\dagger} \; \Delta_j} = \aver{b_i^{\dagger} f_i^{\dagger}
  f_j b_j}$, where the ``pairing'' operator is described by $\Delta_j=f_j
b_j$. Their values have been plotted in Fig.~\ref{fig:Corr} for the same sample
points as before. 
Fourier transforms of bosonic and fermionic Green functions give respective
momentum distributions, while Fourier transform of $G_{bf} (i,j)$ is related
to momentum distribution of composite particles. Such momentum distributions
are shown in Fig.~\ref{fig:NkCorrPaired}. In order to calculate them, it is
crucial to choose the properly defined free-particle eigenmodes due to open
boundaries $\phi_k(j) = \sqrt{2/(L+1)} \; \sin (k \; j)$ with	  $k = n
\frac{\pi}{L+1} (n=1, ... ,L)$. In addition to these  correlation functions,
density-density correlations  $D_{\alpha,\beta} = \aver{n_{\alpha}(i)
  n_{\beta}(j)} - \aver{n_{\alpha}(i)} \, \aver{n_{\beta}(j)} $ were measured
as well. Here $\alpha$ and $ \beta $ can take any value from $\{ b, f\}.$ For
finite size simulations, subtraction of the non-connected part is necessary,
since  open boundary conditions give density profiles with Friedel
oscillations. Examining of density correlations without this substraction
could lead to misinterpretation of results. 

To extract the long-range behavior of correlation functions, we restricted the
 analysis to the regions far away from the boundaries. In order to check
 whether a certain correlation function  has a power-law or exponential decay,
 it is sufficient to test the simplest power-law and exponential forms
\begin{equation}\label{eq:fits}
   G_* (x) = \left.
   \begin{array}{rcl}
      A_* \ e^{-x / d_*} \\
      B_* \ \abs{x}^{-\alpha_*}
   \end{array}
   \right\}
   \ \ \sin (\omega_* x + \varphi_*),
\end{equation}
where the oscillating term is absent for the pure bosonic Green function. The
fermionic oscillation frequency is correctly given by the fermionic density
$\omega_f = \pi \nu_f \simeq \pi / 4$. The exact value is given by the density
in the system bulk, which is slightly larger due to open boundary
conditions. For convenience, let us introduce the Luttinger parameters $K_b,
K_f$ and $K_{bf},$ which are related to $\alpha_b, \alpha_f$ and $\alpha_{bf}$
in Eq.(\ref{eq:fits}) as $\alpha_b=1/(2K_{b}),
\alpha_{f(bf)}=(K_{f(bf)}+1/K_{f(bf)})/2$. Equation~(\ref{eq:fits}) gives the
asymptotic form of the correlation functions in the thermodynamic limit far
from the boundaries. To quantitatively extract the Luttinger parameters $K_*$
for finite $L$, one has to take into account carefully the effects of  open
boundary conditions (OBCs). We refer the reader to the detailed analysis of
Ref.~\cite{Caza04} and recall here that $G_*(i,j)$ will depend on chord
functions $d_L (x) = d\left( x|2(L+1) \right) = \left(2(L+1)/\pi \right)
\, \sin \left( \pi x / 2(L+1) \right)$ of
all $2i, 2j, i-j, i+j$. The form of the correlation function which needs to be
fitted to extract Luttinger parameters are given by ($G_{bf}$ is modified in
the same way as $G_{f}$) 
\begin{eqnarray}
 G_b(i,j) & \propto & \left[ d_L(2i) d_L(2j)
  \right]^{\frac{1}{4 K_b}} \ \left[ d_L(i+j)
  d_L(i-j) \right]^{-\frac{1}{2 K_b}},
\\
G_f(i,j) & \propto & \left[ d_L(2i) d_L(2j)
  \right]^{-\frac{1}{4}(K_f-\frac{1}{K_f})} \ \left[ d_L(i+j)
  d_L(i-j) \right]^{-\frac{1}{2}(K_f+\frac{1}{K_f})}\\
\nonumber  &	  & \left\{ A_0 \, [d_L(i+j)]^{K_f} \, (-1)^{{\rm sign}(i-j)}
  \, \sin \left( \pi \omega (i-j) + \varphi_0 \right) + A_1 \, [d_L(i-j)]^{K_f}
  \, \sin \left( \pi \omega (i+j) + \varphi_1 \right) \right\}.
\end{eqnarray}

We start our tour around the phase diagram by looking at small attractions
between bosons and fermions. The sample point $\gamma_{bb}=6.0$ and
$\gamma_{bf} = -1.36$ is considered in the first panels of
Figs.~\ref{fig:Pop} and \ref{fig:Corr}. Looking at density profiles in
Fig.~\ref{fig:Pop}, one can notice that both the atomic species spread out
around the whole lattice, exhibiting Friedel oscillations due to hard walls at
the boundaries. Wings are cut off for  the sake of plot clearness. Due to
attraction between bosons and fermions  such oscillations are in-phase, but
the values of the two densities differ on the order of the second
digit. Coming to correlation functions (illustrated in Fig.~\ref{fig:Corr}, first
panel), one can easily  recognize a power-law decay for both the bosonic Green
function $G_b (i,j)$ (black circles) and the (oscillating) fermionic  Green
function $G_f (i,j)$  (red squares). ``Pairing'' correlation function $G_{bf}
(i,j)$ exhibits power-law behavior as well. Thus this phase is a two-component
Luttinger liquid. Such phase has two gapless sound modes, and all correlation
functions have algebraic decay.

Increasing the interspecies attraction for small $\gamma_{bb}$ will lead to a
collapse. More precisely, bosons form a small region with high density where
fermions will be attracted up to Pauli-allowed density $\nu = 1.$ Existence of
such maximal density is an artifact of our lattice discretization, and is not
expected in the absence of a lattice. According to mean-field theory
~\cite{Das}, the first order phase transition between the two component LL and
the collapsed phase should  take place as boson-fermion attraction is
increased for any value of the boson-boson interaction. Within mean-field
theory transition line is given by $\gamma_{bf}^2=\gamma_{bb} \pi^2,$ and it
is shown in Fig.~\ref{fig:phasediag} as a solid line. The result of mean field
calculation agrees well with the data set for smallest $\gamma_{bb}$
considered. For large $\gamma_{bb},$ mean-field calculation is not expected to
give an adequate description of the system, and	 for sufficiently large
$\gamma_{bb}$ and attractive fermion-boson term $\gamma_{bf} < 0$ the system
belongs to a third intermediate phase, see Fig.~\ref{fig:phasediag} for a
sketch. The population distributions and the correlations for a sample point
in this intermediate region of parameters are plotted in the second panels of
Figs.~\ref{fig:Pop} and~\ref{fig:Corr}. A strong locking of one density profile on
top of the other is the most striking feature in Fig.~\ref{fig:Pop}(b). Indeed,
not only the Friedel oscillations are in phase as they were in
Fig.~\ref{fig:Pop}(a), but the difference between boson and fermion local
densities is bounded to be less than $10^{-4}$ in the bulk, which is two
orders of magnitude smaller than in the case of two-component LL. Furthermore,
the strong locking of the two densities suggest that a composite particle made
by a boson and a fermion, polaron, could be the new elementary object to  
look at. In the second panel of Fig.~\ref{fig:Corr}  all three types of
correlation functions are plotted: $G_b$ (black circles), $G_f$ (red squares)
and $G_{bf}$ (green diamonds). In contrast to two-component LL, single species
Green functions clearly exhibit an exponential decay with a correlation length
of few sites. However, the ``pairing'' correlations $G_{bf}$ still decay slowly as
a power-law. Taking the open boundary conditions into account as described
before, we get a Luttinger parameter $K_{bf} = 0.95 \pm 0.02$. As shown in
Fig.~\ref{fig:NkCorrPaired}, such a dramatic change in the decay properties of
correlation functions is witnessed by momentum distributions of the two
species (and the composite one). Indeed, the Fermi step of individual
fermionic atoms is no more there as it is in the case of two-component LL, and
also the once peaked Bose distribution is considerably spread out now. In
contrast, we highlight that momentum distribution of paired composite
particles clearly exhibit a Fermi step around $k_{bf} = \pi /4$ consistently
with the filling. Thus, this phase can be understood as the \virg{paired}
phase of bosons and fermions. The existence of such a ``paired'' phase has been
predicted in Ref.~\cite{CazalillaHo} based on LL theory arguments for mixture
with equal masses for large $\gamma_{bb}.$ Indications of the existence of
such phases have also been briefly presented in Ref.~\cite{Pollet06}, but the
phase diagram has not been studied in detail. Figure~\ref{fig:phasediag} presents
the phase diagram for the K-Rb mixture, and shows that ``paired'' phase can be
realized for moderately strong Bose-Bose interactions. Boundaries between
different phases were determined comparing algebraic and exponential fits of
single species correlation functions, and by observing the ``locking'' of one
density profile on the top of the other. 

 Looking at density-density correlations, one can address another distinctive
 feature of ``paired'' phase. As predicted in Ref.~\cite{CazalillaHo},
 oscillating part of all three density correlation functions $D_{\alpha,
   \beta} (x)$  ($b-b$, $f-f$, $b-f$) decay with distance with the same
 algebraic exponent 
\begin{equation}\label{eq:densdens}
 D_* (x)|_{2 \pi \nu} \sim \abs{x}^{-r} \ \sin (\omega x +\varphi) .
\end{equation}
The frequency of oscillations is twice the particle density
 $\omega \simeq 2 \pi \nu \simeq \pi/2$.
 As pointed out by Ref.~\cite{CazalillaHo}, exponent $r$ should be intimately
 related to the Luttinger parameter for paired particles  $K_{bf}$, i.e. $r= 2
 K_{bf}$ (we note that in our notations $K_{bf}=K_+/2,$ where $K_+$ in defined
 in Ref.~\cite{CazalillaHo}).
 Thus parameter $K_{bf}$ can be extracted independently from $D_{\alpha,
   \beta},$ using the fitting procedure which takes OBC into account. 
 We checked that all density-density correlations decay with the same
 exponent, and extracted value of $K_{bf}$ equals $0.97 \pm 0.02.$  This is in
 good agreement with the pairing correlation fits (see before) which give the
 value $0.95 \pm 0.02$. Both these derivations have been checked for scaling
 with respect to both the system size $L$ (from $48$ to $128$) and the DMRG
 truncation parameter $m$ (up to $320$, discarded down to $< 10^{-8}$). The
 values of $K_{bf}$ has been thus confirmed to survive to the thermodynamical
 limit. Based on all evidence, we can unambiguously state that we have shown
 the existence of the \virg{paired} phase predicted by Cazalilla and
 Ho~\cite{CazalillaHo}, even  with moderate interactions and unequal masses of
 the two atomic species for the experimentally relevant case of K-Rb
 mixture. Such a phase survives up to the thermodynamical limit  with an almost
 unchanged exponent near $1$. We recall here that the universal prediction
 of unity value holds only at the transition point itself which, on the other
 hand, is difficult to be precisely addressed due to its BKT nature. 

\begin{figure}[!t]
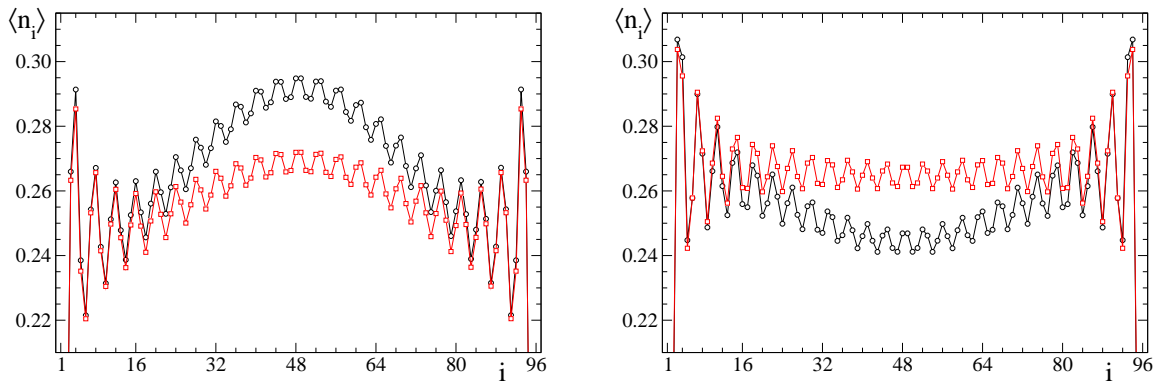

 \begin{center}
\tabcolsep=0.025\linewidth
\begin{tabular}{cc}
\includegraphics[width=0.4\linewidth]{Stab.BP1F0.eps} &
\includegraphics[width=0.4\linewidth]{Stab.B0FP1.eps}
\end{tabular}
\caption{\small{Density profiles in the presence of an extra boson 
(first panel) or fermion (second one) with respect to filling $1/4$ in 
the paired region (same settings as Fig.~\ref{fig:Pop}). The two 
species' profiles retain their ``pairing'' in the wings and exhibit a 
peak or a hole in the bosonic density in the center of the box. For 
larger density imbalance, the system becomes unstable to collapse.} 
\label{fig:stability} }
\end{center}
\end{figure}

The stability of ``paired'' phase against the population imbalance between the
two species was also briefly studied (see Fig.~\ref{fig:stability}). It turns
out that for small enough density differences, the ``locking'' of the densities
survives in the sides of the box, whereas in the middle a peak or a hole
arises in the bosonic profile. The system phase separates into the region
which exhibits ``pairing'' and the region with unequal densities. For larger
density imbalance, the system becomes unstable to collapse, and the ``paired''
phase is washed out. This example qualitatively	 illustrates that the ``paired''
phase can be observed even if the densities of bosons and fermions are not
exactly equal, but the imbalance is smaller than some threshold. More detailed
studies of phases with unequal densities lie beyond the scope of this work.

Let us now concentrate on possible experimental techniques to detect the
``paired'' phase. One notable feature, which distinguishes the ``paired'' phase
from the two-component LL phase, is the presence of the gap for out-of-phase
density modes. Deep in the ``paired'' phase, the energy scale for the gap is
set by Fermi energy $\sim \pi^2\hbar^2n^2/(2m).$ The presence and the size of
the gap can be measured using RF spectroscopy~\cite{RFChin,RFEsslinger}.
Energies of sound modes can be measured using Bragg
scattering~\cite{Bragg_scatt},
and the ``paired'' phase has only one sound mode,
as opposed to two-component LL phase, which has two modes. Disappearance of
the out-of-phase sound mode also affects qualitatively the response of the
system to the variations of the external potential, since out-of-phase
collective mode in external trap is much higher in energy compared to in-phase
mode. Fourier transforms of bosonic and fermionic correlation functions,
$\langle n_b(k)\rangle =\int \phi_k(x) G_b(x) dx$ and $\langle
n_f(k)\rangle=\int \phi_k(x) G_f(x) dx$, are given by momentum distributions,
shown in Fig.~\ref{fig:NkCorrPaired}. They can be measured using ballistic
time-of-flight experiments, since during ballistic expansion momentum
distributions get mapped into real space densities~\cite{Paredes}. The momentum
distribution of composite particles $\langle n_{bf}(k)\rangle=\int \phi_k(x)
\langle\Psi_b^\dagger(x)\Psi_f^\dagger(x)\Psi_f(0)\Psi_b(0)\rangle dx$ has a
strong Fermi step, and can be written as $\langle n_{bf}(k)\rangle=\int
\langle n_b(k-p) n_f(p)\rangle dp/(2\pi).$ We note that it is different from
$\int \langle n_b(k-p)\rangle \langle n_f(p)\rangle dp/(2\pi),$ thus the
presence of the ``pairing'' results in nontrivial noise correlations in
time-of-flight images~\cite{Altman_PRA}. Finally, we would also like to point
out the method to measure the correlation functions based on interference of
two independent 1D clouds~\cite{pnas,fcslong}. For bosons (fermions) the average
of the square of interference signal $\langle|A_{b(f)}(L)|^2\rangle$ of two
segments of length $L$ is related to an integral of the Green's function as
$\langle|A_{b(f)}(L)|^2\rangle=\int_0^L\int_0^L dx dy G_{b(f)}(x-y)^2.$
The interference signal appears at wave vectors $Q_{b(f)},$ which depend on masses
of interfering particles. If one measures $\langle|A_{b(f)}(L)|^2\rangle$ as a
function of $L,$ then in principle dependence of $ G_{b(f)}(x)$ on distance
$x$ can be extracted, since $G_{b(f)}(L)^2=\frac{1}{2}\frac{\partial^2
  \langle|A_{b(f)}(L)|^2\rangle}{\partial L^2}$. The same technique can be used
also to measure $G_{bf}(x),$ but in this case the information will be
contained in the oscillations of the product of Bose and Fermi densities at
wave vector $Q_b + Q_f.$ Since in ``paired'' phase $G_{bf}(x)$ decays much
slower with distance than $G_{b}(x)G_{f}(x),$ ``paired'' phase will be
characterized by strong correlations in the fluctuations of bosonic and
fermionic interference fringes. 

To summarize, we have considered one-dimensional Bose-Fermi mixture with equal
densities and unequal masses using DMRG. For the mass ratio of K-Rb mixture
and attraction between bosons and fermions, we determined the phase diagram,
which is shown in Fig.~\ref{fig:phasediag}. For weak boson-boson interactions,
there is a direct transition between two-component Luttinger liquid and
collapsed phases as the boson-fermion attraction is increased. For strong
enough boson-boson interactions, we find an intermediate ``paired'' phase,
which is a single-component Luttinger liquid of composite particles. We
investigated correlation functions of such a ``paired'' phase, studied its 
stability to density imbalance, and discussed various
experimental techniques which can be used to detect it.

We thank E. Demler, V. Gritsev, R. Fazio, and F. Dolcini for useful 
discussions. This work has been developed using the DMRG code released 
within the "Powder with Power" project (www.qti.sns.it)



\begin{thebibliography}{99}
%
\bibitem{Feshbach}
S.~L. Cornish, {\it et al.}, Phys. Rev. Lett. {\bf 85}, 1795 (2000);
S. Inouye, {\it et al.}, Phys. Rev. Lett. {\bf 93},  183201 (2004);
F. Ferlaino, {\it et al.}, Phys. Rev. A {\bf 73}, 040702(R) (2006);
C.~A. Stan, {\it et al.}, Phys. Rev. Lett. {\bf 93}, 143001  (2004). 
%
\bibitem{Jaksch98}
D. Jaksch, {\it et al.}, Phys. Rev. Lett. {\bf 81}, 3108 (1998).
%
\bibitem{Bloch}
M. Greiner, {\it et al.}, Nature {\bf 415}, 39(2002).
%
\bibitem{Weiss}
T. Kinoshita, T. Wenger and D.S. Weiss, Science {\bf 305}, 1125 (2004). 
%
\bibitem{Paredes} B. Paredes {\it et al.}, Nature {\bf 429}, 277 (2004).
%
\bibitem{bfexp}
B. DeMarco and D.S. Jin, Science {\bf 285}, 1703(1999);
F. Schreck {\it et al.}, Phys. Rev. Lett. {\bf 87}, 080403 (2001);
A.G. Truscott {\it et al.}, Science {\bf 291}, 2570(2001);
G. Modugno {\it et al.}, Science {\bf 297}, 2240 (2002);
Z. Hadzibabic {\it et al.}, Phys. Rev. Lett. {\bf 88}, 160401 (2002);
J. Goldwin {\it et al.}, Phys. Rev. A {\bf 70}, 021601(R) (2004);
G. Roati, {\it et al.}, Phys. Rev. Lett. {\bf 89}, 150403 (2002);
C. Silber, {\it et al.}, Phys. Rev. Lett. {\bf 95}, 170408 (2005).
%
\bibitem{Das} K.K. Das,	 Phys. Rev. Lett. {\bf 90}, 170403 (2003).
%
\bibitem{CazalillaHo}
M. A. Cazalilla, and A. F. Ho, Phys. Rev. Lett. {\bf 91}, 150403 (2003).
%
\bibitem{Mathey}
L. Mathey, {\it et al.}, Phys. Rev. Lett. {\bf 93}, 120404 (2004).
%
\bibitem{MatheyWang}
L. Mathey, D.W. Wang, Phys. Rev. A {\bf 75}, 013612 (2007).
%
\bibitem{Mathey_commensurate}
L. Mathey, Phys. Rev. B {\bf 75}, 144510 (2007).
%
\bibitem{Tmatrix}
X. Barillier-Pertuisel, {\it et al.}, Phys. Rev. A {\bf 77}, 012115 (2008).
%
\bibitem{exactsolution}
A. Imambekov and E.Demler, Phys. Rev. A {\bf 73}, 021602(R) (2006)
 and Ann. Phys. {\bf 321}, 2390 (2006);
M.T. Batchelor, {\it et al.}, Phys. Rev. A {\bf 72}, 061603 (2005);
H. Frahm and G. Palacios, Phys. Rev. A {\bf 72}, 061604(R) (2005).
%
\bibitem{jap_numerics}
Y. Takeuchi and H. Mori, Phys. Rev. A {\bf 72}, 063617 (2005)
and Int. J. Mod. Phys. B {\bf 20}, 617 (2006)
and J. Phys. Soc. Jap. {\bf 74} 3391 (2005). 
%
\bibitem{Sengupta05}
P. Sengupta, and L.P. Pryadko, Phys. Rev. B {\bf 75}, 132507 (2007). 
%
\bibitem{Pollet05}
L. Pollet, {\it et al.}, Phys. Rev. Lett. {\bf 96}, 190402 (2006).
%
\bibitem{Pollet06} L. Pollet {\it et al.}, cond-mat/0609604.
%
\bibitem{currBFM}
K. G\"unter, {\it et al.}, Phys. Rev. Lett. {\bf 96}, 180402 (2006);
S. Ospelkaus {\it et al.}, Phys. Rev. Lett. {\bf 96}, 180403 (2006); 
D.~B.~M. Dickerscheid, {\it et al.}, Phys. Rev. Lett. {\bf 94},  230404
(2005).
%
\bibitem{Kuklov03}
B. Kuklov, and B.V. Svistunov, Phys. Rev. Lett. {\bf 90}, 100401 (2003).
%
\bibitem{Olshanii98} M.Olshanii, Phys. Rev. Lett. {\bf 81}, 938(1998).
%
\bibitem{LL}
E.H. Lieb and W. Liniger, Phys. Rev. {\bf 130}, 1605 (1963);
E.H. Lieb, {\it ibid.}	{\bf 130}, 1616 (1963).
%
\bibitem{SchmidtCazalilla}
M. A. Cazalilla, Phys. Rev. A {\bf 67}, 053606 (2003);
B. Schmidt, L.I. Plimak, and M. Fleischhauer, Phys. Rev. A {\bf 71},
 041601(R) (2005).
%
\bibitem{DMRG}
  S.R. White, Phys. Rev. Lett. {\bf 69}, 2863 (1992);
  S.R. White, Phys. Rev. B {\bf 48}, 10345 (1993);
  U. Schollw\"ock, Rev. Mod. Phys. {\bf 77}, 259 (2005);
  G. De Chiara, M. Rizzi, D. Rossini, S. Montangero, cond-mat/0603842,
  {\it to be published on J. Comput. Theor. Nanosci.}
%
\bibitem{Caza04} M. A. Cazalilla, Journal of Physics B:	AMOP {\bf 37}, S1-S47 (2004).
%
\bibitem{RFChin} C. Chin {\it et al.}, Science {\bf 305}, 1128 (2004).
%
\bibitem{RFEsslinger}
H. Moritz, {\it et al.}, Phys. Rev. Lett. {\bf 94}, 210401 (2005).
%
\bibitem{Bragg_scatt}
D.M. Stamper-Kurn {\it et al.}, Phys.Rev.Lett. {\bf 83}, 2876 (1999). 
%
\bibitem{Altman_PRA}
E. Altman, E. Demler, M. D. Lukin, Phys. Rev. A {\bf 70}, 013603 (2004). 
%
\bibitem{pnas}
A. Polkovnikov, E. Altman and E. Demler, Proc. Natl. Acad. Sci. USA {\bf 103},
6125 (2006).
%
\bibitem{fcslong}
A. Imambekov, V. Gritsev, E. Demler, arXiv:cond-mat/0703766v1,
{\it Proceedings of "Ultracold Fermi gases" Summer School, Varenna, Italy, June 2006}.
%
\end{thebibliography}
\end{document}